\documentclass{amsart}
\usepackage{graphicx}
\newtheorem{thm}{Theorem}[section]
\newtheorem{cor}[thm]{Corollary}
\newtheorem{con}[thm]{Conjecture}
\newtheorem{lem}[thm]{Lemma}
\newtheorem{prop}[thm]{Proposition}
\theoremstyle{definition}
\newtheorem{defn}[thm]{Definition}
\theoremstyle{remark}
\newtheorem{rem}[thm]{Remark}
\numberwithin{equation}{section}

\newcommand{\MA}{\textit{\textbf{A}}}
\newcommand{\MQ}{\textit{\textbf{Q}}}
\newcommand{\Mv}{\textit{\textbf{v}}}
\newcommand{\MD}{\textit{\textbf{D}}}
\newcommand{\MX}{\textit{\textbf{X}}}
\newcommand{\MM}{\textit{\textbf{M}}}
\newcommand{\MU}{\textit{\textbf{U}}}
\newcommand{\MV}{\textit{\textbf{V}}}
\begin{document}

\title[]{Equipectrality and Transplantation}%
\author{Mikl\'{o}s Antal and Mih\'{a}ly Makai}%
\address{BME Institute of Nuclear Techniques, H-1111 Budapest, M\"uegyetem
rkp. 9\\
KFKI Atomic Energy Research Institute, H-1525 Budapest 114, POB49,
Hungary
}%
\email{makai@reak.bme.hu}%

\thanks{}%

\begin{abstract}
We present a technique novel in numerical methods. It compiles the domain of the
numerical methods as a discretized volume. Congruent elements are glued together to
compile the domain over which the solution of a boundary value problem of a linear
operator is sought. We associate a group and a graph to that volume. When the group is
symmetry of the boundary value problem under investigation, one can specify the
structure of the solution, and find out if there are equispectral volumes of a given
type. We show that similarity of the so called auxiliary matrices is sufficient and
necessary for two discretized volumes to be equispectral. A simple example demonstrates
the feasibility of the suggested method.
\end{abstract}
\maketitle
\section{Problem description}
In both science and engineering, we solve boundary values in a
volume composed of large number of meshes. This is the case in
nuclear engineering \cite{AszTo}, in fluid dynamics \cite{BCW},
\cite{CBS} and electromagnetic fields \cite{HS}. We address the
question: under what conditions are the solutions for two meshes
identical, or, are the solutions transformable into each other by a
simple rule? How to find the transformation rule? How can we find
equivalent meshes?
\par
In the design and safety analysis of large industrial devices,
calculational models are tested against experiments carried out on a
small scale mock-up. This is the case with nuclear power plants
\cite{Mah}, aeroplanes \cite{Gupta}, and ships \cite{Vlah}. We would
need a transplantation of the measured values to the geometry of the
real scale device. Is there any hope of doing that exactly or have
we to put up with approximate methods \cite{Cho}?
\par
In the sequel, we assume operator $\mathbf{A}$ to be a linear
operator defined in a finite domain $V$ in $\mathbb{R}^2$ and to
commute with the symmetry group of the plane $\mathbb{R}^2$. As to
its physical meaning, the authors had in their minds the Laplace
operator occurring in several physical problems from electricity to
quantum physics, the diffusion and/or transport operators as used in
reactor physics with homogeneous material distributions. The
boundary value problem considered in the sequel is of Cauchy type
(solution is zero along the boundary of $V$), but the authors are
convinced that generalization to Neumann or third type boundary
value problem is straightforward.
\par
We suggest using \textit{discretized volumes}. The domain $V$, over
which the solution of the boundary value problem is sought is
constructed by the following procedure. We choose an appropriate
simply connected tile $t$, which in our cases is an $n$-gon and glue
copies of the tile by their corresponding sides. Although in
principle $t$ is almost arbitrary \cite{Brooks}, we confine the
discourse to triangular shaped tiles. Since triangulation is a well
known and widely used technique even in theoretical problems
\cite{Still}, this is not considered as a limitation. The
discretized volumes considered by us are always finite. A concise
description of the structure of a discretized volume $V$ is a
finitely presented group {\textbf{G}} \cite{Hamm}. Although in
practical problems the applications of computational group theory
are rather limited, the authors strongly hope for a steady
development in both computational tools (software) and means
(hardware). So, a discretized volume is described by the tile $t$
and the group \textbf{G}. A further asset, a graph $\Gamma$ is also
defined. If copies $t_i$ and $t_j$ of tile $t$ are interconnected by
an edge $\alpha $ of $t$, graph vertices $i$ and $j$ are also
interconnected by an edge $s_{\alpha }$. Analyzing the group
\textbf{G}, the graph $\Gamma$, one can easily reveal basic
properties of $V$. The main results of the present work are:
\begin{enumerate}
\item We provide an algebraic description of the discretized volume. By analyzing the
group and graph associated with a given discretized volume, one can
answer a number of questions. \item Using that algebraic
description, we can formulate conditions for equispectral volumes to
exist. \item We formulate a formal solution to the eigenvalue
problem to be specified later. \item We give conditions for two
discretized volumes to be equipectral and we show that the
eigenfunctions of equispectral discretized volumes are transformed
into each other by a linear map.
\end{enumerate}
\par
The structure of the manuscript is as follows. We define the
discretized volume (DiV) in Section 2, along with the associated
group theoretic assets, group and graph. In Section 3, we present
the solution space $\mathbb{L}_V$, the space of square integrable
functions. By the discretized volume being glued copies of a tile
$t$, the solution is decomposed into functions defined along $t$.
This allows for the dot product to be applied to functions over
different discretized volumes.  Section 4 discusses the structure of
the solution for the boundary value problem under consideration.
Section 5 treats a well known example by the tools presented in
parts preceding Section 4 of the manuscript.
\section{Algebraic Description of Discretized Volumes (DiV)}
In the present section, we study special planar domains over which
we are going to solve eigenvalue problems.
Throughout the present work we investigate cases where $t\subset
\mathbb{R}^2$ is an acute scalene triangle (see Figs. 1, 2).
\begin{defn}\label{D2.2}
A discretized volume $V$ is composed of $N<\infty $ (finite number)
copies of tile $t$ so that we glue copies of tile $t$ to each other
along a corresponding side. Those copies of $t$, which share a side
are called adjacent. The shared side is called internal. When $N>1$,
every copy of $t$ has at least one internal side, see Figs. 1, 2.
\end{defn}
\begin{defn}\label{D2.3}
We say that the discretized volumes $V_{1}$, $V_{2}$ are equivalent
if there is  an isometry of the Euclidean plane $\mathbb{R}^2$ which
maps $V_{1}$ into $V_{2}$. Equivalent discretized volumes are
denoted as $V_{1}\sim V_{2}$.
\end{defn}
We only remark here, that definition \ref{D2.3} does not distinguish
the "warped propellers" \cite{BCD} because they are obtained from
each other by interchanging the dotted and dashed sides of $t$ and
that operation leaves $t$ invariant.
Our method for constructing DiVs has immanent limitations. Depending
on the tile $t$, we may tile out the entire plane, or, after a given
number of gluing, we have to stop because the next glued copy would
intersect with an already existing copy. To clear that problem, we
investigate the transformation rules of tile $t$ under gluing.
\begin{defn}\label{Glu-1}
We label the corresponding side of the congruent copies of tile $t$
in DiV $V$ by $\alpha, \beta, \gamma$. Let $\mathbf{T}_\nu$ be a
local reflection on the side $\nu$ of $t$, $\nu\in\{\alpha, \beta,
\gamma\}$. The image of $t$ under $\mathbf{T}_\nu$ is denoted by
$t_\nu=\mathbf{T}_\nu\cdot t$.
\end{defn}
By gluing, we get a new copy, $t_\nu$, and gluing is applicable
to the new copy as well, so we can define an operation among the
$\mathbf{T}_\nu$ transformations and that operation is the
consecutive application of gluing.
\begin{defn}\label{Glu-2}
Let $\mathbf{T}_\nu=\mathbf{T}_{\nu_1}\mathbf{T}_{\nu_2}$ mean the
following transformation: apply first $\mathbf{T}_{\nu_2}$ and then
$\mathbf{T}_{\nu_1}$ to the result:
$\mathbf{T}_{\nu_1}\mathbf{T}_{\nu_2}t=\mathbf{T}_{\nu_1}\left(\mathbf{T}_{\nu_2}\right)t$,
see Figs. 1, 2.
\end{defn}
We are going to use the term {\em set of gluing} to the set of
transformations defined above. On the set of transformations we
defined multiplication, and there is a unit element $\mathbf{T}_e$
which leaves $t$ invariant. Such an element is, for example, the
repeated application of $\mathbf{T}_\nu$. Trivially there is also an
inverse, because if $t_\nu$ is obtained by gluing from $t$ then also
$t$ is obtained from $t_\nu$ by gluing.
\begin{defn}\label{Glu-3}
The set of transformations
$\{\mathbf{T}_{\nu_1}\mathbf{T}_{\nu_2}\dots\mathbf{T}_{\nu_n}\}$ is
endowed with the multiplication, see definition \ref{Glu-2}, there
is an identity element $\mathbf{T}_e=\mathbf{T}_\nu \mathbf{T}_\nu$,
and every element has an inverse, therefore the transformations form
a group \textbf{G}$_t$.
\end{defn}
We are going to refer to the group of transformations as
\textbf{G}$_t$, where $t$ refers to the tile, the corner stone of
DiVs. Note that we allowed for a repeated application of a given
transformation and then we get back the original tile. But there are
tiles which after a sequence of transformations only partly coincide
with a former copy of tile $t$. We exclude such a situation from our
investigation.
\begin{defn}\label{Glu-4}
We call
$\mathbf{T}\in$\textbf{G}$_t$ realizable if for any factorization of
$\mathbf{T}=\mathbf{T}_\mu \mathbf{T}_\nu$ the images
$\mathbf{T}_\mu\cdot t$ and $\mathbf{T}_\nu\cdot t$ are either
disjoint or coincide for any $\mathbf{T}\in$ \textbf{G}$_t$.
\end{defn}
The set of realizable transformations depends solely on the tile
$t$. From now on, we deal solely with realizable $\mathbf{T}$
transformations.
\par
Now we pass on to investigate the discretized volume $V$. We wish to associate algebraic
descriptions with $V$, we define a graph and a group.
\begin{defn}\label{D2.4}
A graph $\Gamma_V$ is assigned to $V$ in the following way. We label
the copies of $t$ in $V$. If the copies labeled as $n_{1}$ and
$n_{2}$ are adjacent, and they share a side of type $\alpha $, then
the vertices $n_{1}$ and $n_{2}$ of graph $\Gamma_V$ are connected
by an edge of type $\alpha$.
\end{defn}
\begin{defn}\label{D2.5}
We associate a permutation group \textbf{G} to $V$ in the following
way. When in $V$ side $\alpha $ of type $a$ connects the copies
$t_{i_{\alpha 1}}$, $t_{i_{\alpha 2}}$; $t_{j_{\alpha 1}}$,
$t_{j_{\alpha 2}},{\ldots}$ then, we form the permutation $a= (
i_{\alpha 1}, i_{\alpha 2}),( j_{\alpha 1},j_{\alpha 2}){\ldots}$.
We repeat that procedure for sides $\alpha $, $\beta $, $\gamma $ to
get generators $a, b$, and $c$, and group \textbf{G} is generated by
$a, b$, and $c$.
\end{defn}
The next step i to define the group action of \textbf{G} on $V$, and
$t$, see Fig. 1, and Table 1.
There is a natural map of an element $g\in$\textbf{\textit{G}}  to
an element $\mathbf{T}_\mu \in$\textbf{G}$_t$. Either group is
finitely presented, the number of generators is the same for both
groups. Let \textbf{G}$=<a,b,c|a^2=b^2=c^2=e>$ and
\textbf{G}$_t=<\mathbf{T}_\alpha, \mathbf{T}_\beta,
\mathbf{T}_\gamma|\mathbf{T}^2_\alpha=
{\mathbf{T}_\beta}^2={\mathbf{T}_\gamma}^2=\mathbf{T}_e>$. Then the
map $\alpha=f(a), \beta=f(b), \gamma=f(c)$ maps \textbf{G}$\to$
\textbf{G}$_t$ and is an injection. Such a permutation
representation is a so called faithful representation. We apply the
notation $\nu=f(g)$, $g\in \{a,b,c\}$.
\begin{defn}\label{D2.6}
The action $(g\cdot t)$ of $g\in$ \textbf{G} on a tile $t$ is
defined as $\mathbf{T}_\nu\cdot t$ where $\nu=f(g)$.
\end{defn}
\begin{defn}\label{D2.7}
 The orbit of $t$ under the group $\textbf{G}$ is the set $\mathbf{T}_g\cdot t$  for all
$g\in \textbf{G}$.
\end{defn}
\begin{defn}\label{D2.8}
The action $g\cdot V$ of $g\in \textbf{G}$ on $V$
 is defined as follows. Let $x\in\{a,b,c\}$ be a
generator of $\textbf{G}$. The action of $x$ on copy $t_i$ is $x\cdot t_i=t_i$ whenever
side $f^{-1}(x)$ of copy $t_i$ is not internal in $V$. Otherwise $x\cdot t_i=t_j$ if
copies $t_j$ and $t_i$ $(i,j<N)$ share an internal side of type $f^{-1}(x)$.
\end{defn}
\begin{rem}\label{R1}
With a given tile $t$, $N$ may be limited, see definition
\ref{Glu-4}.
\end{rem}
\begin{defn}\label{D2.9}
Adjacency matrix \MA$_{V}$ of $V$ is an $N\times N$ matrix, its
$a_{ij}$ element is $a_{ij}=1$ if copies $t_i$ and $t_j$ are
adjacent, otherwise $a_{ij}=0$.
\end{defn}
\begin{defn}\label{D2.10}
The auxiliary matrix {\MX} is \MX=\MD + \MA$_V$, where {\MD}  is a
diagonal matrix, its $i^{th }$entry $D_{ii}$ equals the number of
internal sides of copy $t_i$ in $V$.
\end{defn}
Discretized volumes are shown in Fig. 2, with tile $t$, a regular
triangle; with two, three, and four copies of $t$. The side types
are solid (side $a$), dashed (side $b$), and dotted (side $c$)
lines. The discretized volumes are described by $t$ (identity
element), $\{t,\alpha\cdot t\}$, $\{t,\alpha\cdot t,
\beta\cdot\alpha\cdot t\}$, $\{t,\alpha\cdot t,\beta\cdot\alpha\cdot
t, \gamma\cdot\beta\cdot\alpha\cdot t\}$. Let us number the copies
of $t$ as follows: $t\to 1$, $\alpha\cdot t\to 2$,$\beta\cdot
\alpha\cdot t\to 3$, $\gamma\cdot\beta\cdot\alpha\cdot t\to 4$. The
group action on the four elements is given in table \ref{Tab1}.
\begin{table}
  \centering
  \caption{Group action in Fig. 1.}\label{Tab1}
  \begin{tabular}{|c|c|c|c|c|}
    \hline
    generator/copy & 1 & 2 & 3 & 4 \\
    \hline
    a & 2 & 1 & 3 & 4 \\
    b & 1 & 3 & 2 & 4 \\
    c & 1 & 2 & 4 & 3 \\
    \hline
  \end{tabular}
\end{table}
We have elaborated the basic means to be used to analyze the
solution of an eigenvalue problem over a DiV. It is well known that
it suffices to solve an eigenvalue problem over one element of the
set of equivalent discretized volumes. This is because the
equivalence provides a map between equivalent volumes and when that
map is built from transformations commuting with the operator in the
eigenvalue problem we immediately get a transformation of the
solutions. We need further means to recognize if two complex DiVs
are equivalent.
\begin{lem}\label{L2.11}
If $V_{1}\sim V_{2}$ and the corresponding graphs are $\Gamma_1$ and
$\Gamma_2$, then $\Gamma_1\sim \Gamma_2$.
\end{lem}
\textit{Proof:} If the stipulated conditions are met then there is
an isomorphism between the copies and edges of the two discretized
volumes this entails the statement. $\Box$
\begin{lem}\label{L2.12}
Let the number of copies of $t$ in discretized volume $V$ be $N$.
Then, in group \emph{\textbf{G}} associated with $V$, there is a
subgroup of index $N$. Consequently, the order of group
\emph{\textbf{G}} is a multiple of the number of copies $N$ in $V$.
\end{lem}
\textit{Proof:} There is a map between finite groups and Cayley
graphs. A coset representation of $\textbf{G}$ is isomorphic with
graph $\Gamma$, in which there are $N$ vertices. $\Box$
\par
In accordance with Lemma \ref{L2.11}, $\textbf{G}_{1}\sim
\textbf{G}_{2}$ suffices for $V_{1}\sim V_{2}$. In accordance with
Lemma \ref{L2.12}, \textbf{G}$_{1} \sim$ \textbf{G}$_{2}$ suffices
for $V_{1} \sim V_{2}$.
\par
The next section deals with functions defined over a discretized volume.
\section{Function space $\mathbb{L}_V$}
The specific structure of the DiV can be exploited in the analysis of eigenfunction
defined on $V$. Let us consider a function{\footnote{We use the notation $f(x)$ as a
general function, not necessarily the same as before.}} $f(x), x\in V$. We trace back
$f(x)$ to $N$-tuples (here $N$ is the number of copies in $V$), which is a vector space.
Thus we can speak of linear independence of two functions or the dimension of $f(x)$. We
set forth the following following notation.
\par
$x=(x_1,x_2)$ denotes a point in the discretized volume $V$. In tile
$t$, we use a local coordinate $\xi=(\xi_1,\xi_2)$. Since $V$ is
composed of copies of $t$, and copy $t_i$ is obtained as
$t_i=\mathbf{T}_i\cdot t$, and transformation $t\to t_i$ is the
automorphism of the plane, in other words a member of the Euclidian
group $E(2)$, the following coordinate transformation, which acts on
triples $(x_1,x_2,1)$ and is associated with $T_i$:
\begin{equation}\label{Fun-1}
    g_i(\theta_i,a_i,b_i)=\left(%
\begin{array}{ccc}
  \cos \theta_i & \sin\theta_i & 0 \\
  \sin\theta_i & \cos\theta_i & 0 \\
  a_i & b_i & 1 \\
\end{array}%
\right).
\end{equation}
That transformation maps a point $x=(x_1,x_2)$ into
\begin{equation}\label{Fun-2}
    xg_i=(x_1\cos\theta_i + x_2\sin\theta_i+a_i,-x_1\sin\theta_i +x_2\cos\theta_i+b_i).
\end{equation}
The map has the property $xg_ig_j=(xg_i)g_j, g_i,g_j\in E(2)$. Action of $g_i$ on a
function $f(x)$ is
\begin{equation}\label{Fun-3}
    \mathbf{T}_if(x)=f(xg_i).
\end{equation}
Using the above definitions, we have a transformation $t_i\to t$ as
\begin{equation}\label{Fun-4}
    \xi=xg_i^{-1}.
\end{equation}
For a given $x\in V$ also holds $x\in t_i$ for some $1\le i\le N$,
i.e. point $x$ belongs to one of the copies of tile $t$. This gives
rise to a map $f(x)$ to $f(\xi)$ a value at point $\xi$ of $t$:
\begin{equation}\label{Fun-5}
    (\mathbf{T}_i)^{-1}f(x)=f_i(\xi), x\in t_i; \xi\in t.
\end{equation}
When $V$ is composed of $N$ copies of tile $t$, any $f(x)$ is
exhaustively described by the $N$-tuple of functions
$\underline{f}(\xi)=(f_1(\xi), f_2(\xi),\dots,f_N(\xi))$, where
$f_i(\xi)=(\mathbf{T}_i)^{-1}f(x), x\in t_i$.
\begin{defn}
The $N$-tuple $\underline{f}=(f_1(\xi), f_2(\xi),\dots,f_N(\xi))$, where
$f_i(\xi)=(\mathbf{T}_i)^{-1}f(x), x\in t_i$ is called the vector form of function
$f(x)$ defined over the discretized volume $V$.
\end{defn}
\begin{defn}[Function space $\mathbb{L}_V$]
The function space $\mathbb{L}_V$ contains square integrable functions over discretized
volume $V$.
\end{defn}
\begin{defn}[Dot product and norm]\label{Dot}
Let $f,h\in \mathbb{L}_V$. Then the dot product of $f$ and $h$ is
\begin{equation}\label{Fun-6}
    (f,h)_V=\sum_{i=1}^N \int_t f_i(\xi)h_i(\xi)d\xi,
\end{equation}
where $f_i(\xi)=(\mathbf{T}_i)^{-1}f(x), x\in t_i$. The notation for
$h_i$ is analogous. The norm of $f(x)\in \mathbb{L}_V$ is
\begin{equation}\label{Fun-7}
    ||f(x)||^2=\sum_{i=1}^N \int_t f_i^2(\xi)d\xi.
\end{equation}
\end{defn}
It is evident that the dot product (\ref{Fun-6}) meets the following
general properties of the dot product (here $f, f_1, f_2, h$ are
functions, $a_1$ and $a_2$ are numbers):
\begin{itemize}
  \item symmetry: $$ (f,h)_V=(h,f)_V;$$
  \item Schwartz inequality: $$ (f,h)_V\le ||f|| ||h||;$$
  \item linearity $$(a_1f_1+a_2f_2,h)=a_1(f_1,h)_V + a_2(f_2,h).$$
\end{itemize}
\begin{defn}[Dimension of $f(x)$]
$dim(f(x))$ is the number of linearly independent $f_i(\xi)$
functions in the vector form of $f(x)$. $dim(f(x))$ is called the
dimension of function $f(x)$, defined over the discretized volume
$V$.
\end{defn}
The dot product (\ref{Fun-6}) is actually formulated through the vector forms of the
involved functions:
\begin{equation}\label{Fun-8}
(f(x),h(x))_V=(\underline{f}^+(\xi)\underline{g}(\xi))_t=\sum_{i=1}^N
\int_t f_i(\xi)h_i(\xi)d\xi.
\end{equation}
Since
\begin{equation}\label{Fun-9}
    ({\MM}\underline{f}^+\underline{h})_t= (\underline{f}^+ {\MM}^+
    \underline{h})_t
\end{equation}
for any $N\times N$ matrix {\MM}, we may use the usual rules of
vector dot products:
\begin{enumerate}
    \item ($\underline{f}(\xi),\underline{h}(\xi))_t$=({\MU}$\underline{f}(\xi)$,{\MU}$\underline{h}(\xi))_t$ if
    {\MU} is a unitary matrix.
    \item (${\MM}^+\underline{f},\underline{h})_t=(\underline{f},{\MM}\underline{h})_t$
    \item vectors $\underline{f}$ and $\underline{h}$ are called orthogonal if $(\underline{f},\underline{h})_t=0$, or, what is equivalent,
    if $(f(x),h(x))_V=0$.
\end{enumerate}
Note that solely $N$, the number of copies in the volume of the
integration, and the tile $t$ are relevant in the dot product,
therefore $\underline{f}$ and $\underline{h}$ may belong to
different discretized volumes, provided each of them is composed of
$N$ copies of tile $t$.
\par
 The next section deals with the structure of the solution to the eigenvalue problem.
\subsection{Equispectral discretized volumes} The goal of our investigation is to find
out if there are non-equivalent DiVs that allow for transforming the solutions into each
other. We follow Baron M\"unchausen's procedure{\footnote{Tale hero Baron M\"unchausen
once fell into a swamp and drew out himself by his own forelock.}}: we assume that the
solution is known along the internal boundaries of the DiV and give a formal solution in
term of that. This leads us not only to the structure of the solution but also to a
linear transformation mapping the solutions over two DiVs into each other.
\par
Let us investigate the following eigenvalue problem in a DiV $V_1$:
\begin{equation}
\label{eq1} {\rm {\bf A}}\Phi (x)=\lambda \Phi (x),\,\,x\in V_1.
\end{equation}
We assume $\Phi(x)$ to be in $\mathbb{C}^1(V_1)$, i.e. the function
and its first derivative are continuous in $V$. Furthermore we
assume $\mathbf{A}$ to commute with the automorphisms of $V_1$:
\begin{equation}
\label{eq2} \left[ {{\rm {\bf A}},{\rm {\bf O}}} \right]={\rm {\bf
AO}}-{\rm {\bf OA}}=0.
\end{equation}
Here ${\mathbf{O}}$ is an automorphism of $V_1$, viz. a reflection,
translation or rotation operator acting on functions defined over
$V_1$. Eq. (\ref{eq2}) is the usual definition of symmetries of
operator $\mathbf{A}$. The set of operations with which $V_1$ is
formed involves only symmetries of operator $\mathbf{A}$.
Definitions \ref{D2.6} and \ref{D2.8} assure group $\textbf{G}$,
defined there, to be isomorphic to a group of transformations
commuting with $\mathbf{A}$. It is well known that the solution of
eigenvalue problem (\ref{eq1}) with a homogeneous boundary condition
along the boundary $\partial V_1$ of $V_1$ is easily determined from
a solution of the same problem over another member of the class
$\sim V_1$ (see Definition \ref{D2.3}). The question is, if we can
find a volume $V_2$ not equivalent to $V_1$ such that all the
eigenvalues of problem (\ref{eq1}) will remain the same as for
$V_1$.
\begin{defn}\label{D2.13}
If there exist volumes $V_1$ and $V_2$ each one composed of the same
number of copies of the same tile $t$, so that $V_1$ and $V_2$ are
not equivalent according to definition \ref{D2.3} furthermore all
the eigenvalues of problem (\ref{eq1}) are the same in $V_1$ and
$V_2$, we call $V_1$ and $V_2$ equispectral.
\end{defn}
When speaking about a fixed operator $\mathbf{A}$, its spectrum is
different on different DiVs. We express the formal solution in terms
of assumed known solution along internal boundaries. Can those
functions along an internal boundary be identically zero? Below we
show that either the solution along each internal boundary is
identically zero--that is the degenerate case--or, on every internal
boundaries the solution differs from zero.
\begin{defn}
The set $S$ of eigenvalues $\lambda$ in Eq.(\ref{eq1}) supplemented
with $\Phi(x_0)=0, x_0\in\partial V$ is called the spectrum of
operator $\mathbf{A}$  on DiV $V_1$.
\end{defn}
\begin{lem}
Let $S_{V_1}$ be the spectrum of operator \emph{\textbf{A}} on DiV
$V_1$ composed from tile $t$. Let $S_t$ be the spectrum of operator
\emph{\textbf{A}} on $t$. Let $\lambda_0\in S_t\bigcap S_{V_1}$ and
$\Phi_0(x)$ be the associated eigenfunction in $V_1$. Then
$\Phi_0(x)$ is identically zero on every internal boundary of $V_1$.
\end{lem}
{\em Proof:} See \cite{Her}.$\Box$
\par
The benefit from knowing equispectral volumes comes from the fact
that it is rather tiresome to solve Eq. (\ref{eq1}) even for a
simple volume. At the same time we know that equivalent volumes
provide an easily feasible recipe for transplanting the solution
from one member of the class to another. Knowledge of equispectral
volumes would widen the range of transformations where instead of
solving Eq. (\ref{eq1}) over a new equispectral volume, one would
apply a relatively simple transformation to an already known
solution.
\par
Actually, in a number of practical problems, one would be satisfied
with the equivalence of one eigenvalue. In other words, with the
equality of the respective eigenvalue and a transformation rule for
the eigenfunctions. In a number of cases physical meaning is
attributed to the so called fundamental mode eigenvalue.
\section{Solution of a Boundary Value Problem over DiVs}
\subsection{The formal solution}
Below we derive a formal solution \cite{MaOr} to problem (\ref{eq1})
with Dirichlet boundary condition (i.e. $\Phi =0$ on the boundary).
The solution is given in terms of the Green's function
$\mathcal{G}_{t}$ of tile $t$, that we obtain as the solution of the
following boundary value problem:
\begin{equation}\label{eq3}
    \left({\bf A}-\lambda\right) \mathcal{G}_t (\xi,x_0 )=0, \xi\in t,
\end{equation}
\begin{equation}\label{eq4}
    \mathcal{G}_t (\xi,x_0 )=\delta (\xi-x_0 ), \quad x_0 \in\partial t.
\end{equation}
Our goal is to express the solution of (\ref{eq1}) in terms of given
values along the boundaries. In general, the boundary value uniquely
determines the solution.  In order to build up the solution in DiV
$V$, we build up the solution in a tile $t$ from the values given
along the boundary of $t$. Let the solution of the boundary value
problem
\begin{equation}
\label{eq5} {\rm {\bf A}}\Phi (\xi)=\lambda \Phi (\xi),\xi\in t
\end{equation}
with boundary conditions
\begin{equation}
\label{eq6} \Phi (\xi=x_0)=\left\{ {{\begin{array}{*{20}c}
 {f_a (x_0 ),\,x_0 \in \partial t_a } \hfill \\
 {f_b (x_0 ),\,x_0 \in \partial t_b } \hfill \\
 {f_c (x_0 ),\,x_0 \in \partial t_c } \hfill \\
\end{array} }} \right.,
\end{equation}
where the three sides of tile $t$ are $\partial $t$_{a}$, $\partial
t_{b}$ and $\partial t_{c}$. Then, for $\xi\in t$, we get
\begin{equation}\label{eq7-1}
    \Phi (\xi)=\int\limits_{\partial
t_a } {\mathcal{G}_t (\xi,x_0 )f_a (x_0 )dx_0 +}
\int\limits_{\partial t_b } {\mathcal{G}_t (\xi x_0 )f_b (x_0 )dx_0
+} \int\limits_{\partial t_c } {\mathcal{G}_t (\xi,x_0 )f_c (x_0
)dx_0 } , \xi\in t.
\end{equation}
Since $V$ consists of copies of $t$, the solution of Eq. (\ref{eq1})
in $V$ is the sum of integrals like Eq. (\ref{eq7-1}). The $\lambda$
eigenvalue of operator $\mathbf{A}$ is called degenerate if in the
case $f_a(\xi)=f_b(\xi)=f_c(\xi)=0$ there exists a not identically
zero $\Phi(\xi)$ solution over a tile $t$. The degenerate
eigenfunctions are related to individual solutions over a tile with
zero values fixed on the boundary.
\begin{cor}\label{Cor-1}
For non-degenerate eigenvalues $\lambda$, there is a one-to-one map
between the solution $\Phi(\xi), \xi\in t$ and the conditions
prescribed on the boundary $\partial t$. When the conditions
$(f_{a1}(x_0),f_{b1}(x_0),f_{c1}(x_0))$  and
$(f_{a2}(x_0),f_{b2}(x_0),f_{c2}(x_0))$given along the respective
boundaries of $t_1$ and $t_2$ are linearly independent, the
corresponding solutions $\Phi_1(\xi), \Phi_2(\xi)$ are also linearly
independent.
\end{cor}
Corollary \ref{Cor-1} connects the solution in a DiV $V$ to the
solutions along the boundaries in $V$. When dealing with a Cauchy
boundary value problem, we have only to deal with the internal
boundaries. Let us assume that there are $K$ internal sides (c.f.
Definition \ref{D2.2}) in $V$ and pretend the solution  $f_{k}(\xi
)$ to be known along internal side $k$. We use the vector form of
$\Phi(x)$, in which $N$ functions $\underline{F}(\xi)=(\Phi
_{1}(\xi),{\ldots}, \Phi_{N}(\xi))$, are encountered, one from each
copy of $t$. Then,
\begin{equation}
\label{eq8-1}  \underline{F}(\xi)={\rm {\MQ\Mv}}(\xi),
\end{equation}
where elements of {\Mv}$(\xi)$ are:
\begin{equation}\label{eq9}
    v_{k} (\xi)=\int\limits_{\partial t_k } {\mathcal{G}_t (\xi,x_0 )f_k (x_0
)d x_0 } , k=1,{\ldots},K.
\end{equation}
{\MQ} is an $K\times N$ matrix\footnote{ Since every reflection
creates an internal side, K=N-1.}, and assigns the $K$ internal
sides to the $N$ copies of $t$. As we see, expression (\ref{eq8-1})
has two components, {\Mv}(x) depends only on tile $t$ and operator
{$\mathbf{A}$}, hence we call it the \emph{physical part} of the
solution. From now on, {\Mv},{\Mv}$_1$, {\Mv}$_2,\dots$ denote $K$
tuples, underlined letters denote $N$-tuples. Because of
(\ref{eq3}), and the assumed linearity of $\mathbf{A}$, we get
\begin{equation}\label{eq9-2}
    \mathbf{A}\underline{F}(\xi)={\mathbf{A}\MQ}{\Mv}(\xi)={\MQ\mathbf{A}}{\Mv}(\xi)=\lambda \underline{F}(\xi)
\end{equation}
for any $\xi\in t$. Note that {\Mv}  depends also on the considered
eigenvalue. The formal solution (\ref{eq8-1}) reports us that the
source of any non-identically zero solution derives from at least
one non-identically zero function given along some boundary. We are
dealing with a homogeneous problem therefore the solution is taken
as zero at the external boundary of $V$. When the solution is not
identically zero, we may assume that it differs from zero along the
internal boundaries. In general, each boundary may be independent
and the solution to boundary value problem (\ref{eq5}) must be
linear combination of the solutions coming from the internal
boundaries. Since the Green's function preserves linear dependence
and independence, the dimension of the solution will not exceed the
number of internal sides.
\par
On the other hand, {\MQ} depends only on the structure of $V$, hence
we call it the \emph{structural part} of the solution. Pattern
(\ref{eq8-1}) can be used to derive an integral equation set for the
solution along the internal boundaries and may serve as basis for
approximate solution methods. Now we are interested in the
structural part.
\begin{lem}\label{L3.1}
The auxiliary matrix \MX of volume $V$ and the structural matrix
{\MQ} in Eq. (\ref{eq8-1}) are related as {\MX}={\MQ\MQ}$^{+}$.
\end{lem}
\textit{Proof: }By the proposition, matrix element $X_{ij}$ is the
dot product of the $i^{th}$ and $j^{th}$ rows of matrix {\MQ}.
Hence, $X_{ii}$ equals the number of internal sides of copy i.
Element $X_{ij}$ is 1 if copy i and j share an internal side, and
zero otherwise. This is just the definition of {\MX}. $\Box$
\par
Now we address the question of two discretized volumes $V_1$ and
$V_2$ being equispectral.
\begin{thm}\label{L3.2}
Let discretized volumes volumes $V_1$ and $V_2$ such that
\begin{enumerate}
\item V$_{1}$ and V$_{2}$ are composed of the same tile $t$. \item In V$_{1}$ and
V$_{2}$ the number of copies (N) of $t$ are equal.
\item Along the external boundary of
V$_{1}$ and V$_{2}$ the number of sides are equal by side types.
\item It follows from item 3 that along the
internal boundaries of V$_{1}$ and V$_{2}$ the number of sides are
equal by side types.
\end{enumerate}
Let the formal solutions over $V_1$ and $V_2$ be
$$ \underline{F}_m(\xi)={\MQ}_m{\Mv}_i(\xi), m=1,2.$$
Let the component {\Mv} of the formal solution such that
$$ \mathbf{A} {\Mv}_m=\lambda {\Mv}_m, m=1,2$$
holds. Furthermore, let the auxiliary matrices {\MX}$_1$ and
{\MX}$_2$ be similar. Then, and only then $V_1$ and $V_2$ are
equispectral.
\end{thm}
\textit{Proof: } With appropriate norm, the eigenvalues are given by
\begin{equation}\label{New-1}
\lambda_m=(\mathbf{A}{\MQ}_m{\Mv}_m(\xi);{\MQ}_m{\Mv}_m(\xi)),
m=1,2.
\end{equation}
Under the stipulated conditions  {\MQ}$_1$, {\MQ}$_2$ are $N\times
K$ matrices, where $N$ is the number of copies of tile $t$ in $V_m$,
and $K$ is the number of internal sides in either $V_m$.
\par
First we show that if the auxiliary matrices {\MX}$_1$ and {\MX}$_2$
are similar, then $V_1$ and $V_2$ are equispectral. Using
(\ref{New-1}), we get
$$ \lambda_1-\lambda_2=(\mathbf{A}{\Mv}_1;{\MQ}_1^+{\MQ}_1{\Mv}_1)-
(\mathbf{A}{\Mv}_2;{\MQ}_2^+ {\MQ}_2{\Mv}_2).$$ Because of the
similarity of the auxiliary matrices, we may write
{\MQ}$_2^+${\MQ}$_2$={\MU}{\MQ}$_1^+${\MQ}$_1${\MU}$^+$,
 and ${\Mv}_2={\MU}{\Mv}_1$ where {\MU} is a
unitary $K\times K$ matrix. Now we perform a series of elementary,
identical, and algebraic transformations:
\begin{eqnarray}
 \label{Line1} (\mathbf{A}{\Mv}_2;{\MQ}_2^+{\MQ}_2{\Mv}_2) &=& (\mathbf{A}{\MU}{\Mv}_1;{\MQ}_2^+{\MQ}_2{\MU}{\Mv}_1)= \\
 \label{Line2} ({\MU}\mathbf{A}{\Mv}_1;{\MQ}_2^+ {\MQ}_2{\MU\Mv}_1) &=& ({\MU}\mathbf{A}{\Mv}_1;{\MU}{\MQ}_1^+{\MQ}_1{\MU}^{-1}{\MU}{\Mv}_1)=\\
\label{Line3}  &=& (\mathbf{A}{\Mv}_1;{\MQ}_1^+ {\MQ}_1{\Mv}_1).
\end{eqnarray}
In (\ref{Line1}), we expressed {\Mv}$_2$ by {\Mv}$_1$, in
(\ref{Line2}) we expressed {\MQ}$_2^+${\MQ}$_2$  by
{\MQ}$_1^+${\MQ}$_1$ and exploited that operator $\mathbf{A}$ acts
only on {\Mv}$_i$, and in (\ref{Line3}) we used the properties of
the dot product.
Using the last form, we get
$$\lambda_1-\lambda_2=(\mathbf{A}{\Mv}_1;({\MQ}_1^+{\MQ}_1{\Mv}_1)-{\MQ}_1^+{\MQ}_1{\Mv}_1)=0,$$
from which follows $\lambda_1=\lambda_2$. \par Now we show that if $\lambda_1=\lambda_2$
then the auxiliary matrices are similar. By the assumption, we have
$$ (\mathbf{A}{\Mv}_1;{\MQ}_1^+{\MQ}_1{\Mv}_1)=
(\mathbf{A}{\Mv}_2;{\MQ}_2^+{\MQ}_2{\Mv}_2).$$ From that expression,
operator $\mathbf{A}$ may be left out as it brings in a
multiplication by the same number. Thus we have
$$ ({\Mv}_1;{\MQ}_1^+{\MQ}_1{\Mv}_1)=
({\Mv}_2;{\MQ}_2^+ {\MQ}_2{\Mv}_2).$$ That expression holds for any
components of vectors {\Mv}$_1$ and {\Mv}$_2$ because in the general
case the solution along the $K$ internal faces are independent, and
the Green's function preserve that. Then, the two components of the
dot product may differ only in a rotation: ${\Mv}_2={\MU}{\Mv}_1$
and ${\MQ}_2^+{\MQ}_2 {\Mv}_2={\MU}{\MQ}_1^+ {\MQ}_1{\Mv}_1$. From
this immediately follows ${\MQ}_2^+ {\MQ}_2
={\MU}{\MQ}_1^+{\MQ}_1{\MU}^{-1}$. We have to remember that
{\MQ}$_2^+${\MQ}$_2$ is not the matrix {\MX}$_2$, and from the
similarity of {\MQ}$_2^+${\MQ}$_2$ and {\MQ}$_1^+${\MQ}$_1$ does not
follow immediately the similarity of matrices {\MX}$_2$ and
{\MX}$_1$. We need some matrix theory \cite{Rozsa} to go on.
\par
The $N\times K$ structural matrix {\MQ}$_1$ can be written as
\begin{equation}\label{New-5}
    {\MQ}_1={\MU}{\MD}{\MV}^+
\end{equation}
where {\MU} is an $N\times N$ orthogonal matrix, {\MD} is $N\times
K$ matrix, its first $K$ rows form a diagonal matrix, and the
$K+1,\dots, N$ rows contain
only zeros, {\MV} is an $K\times K$ orthogonal matrix. 
Since
\begin{equation}\label{New-6}
    {\MQ}{\MQ}^+={\MU}{\MD}{\MV}^+{\MV}{\MD}{\MU}^+={\MU}{\MD}{\MD}{\MU}^+,
\end{equation}
it follows that the nonzero eigenvalues of {\MQ}{\MQ}$^+$ and
{\MQ}$^+${\MQ} are the same, therefore {\MX}$_1$ and {\MX}$_2$ are
similar matrices as stated. $\Box$
\par
Let us return to the case of a degenerate eigenvalue. When $\lambda$
is a degenerate eigenvalue, we get a non-zero solution on a tile
with identically zero values prescribed along the boundary of $t$.
With a triangular tile $t$, the copies of $t$ in discretized volume
$V$ fall into two categories (say black and white) as we can color
$V$ by the black and white colors. Let $\underline{w}$ be an $N$
tuple with elements +1 or -1 in position $i$ when copy $t_i$ is
black or white, respectively.
\begin{lem}\label{L3.4}
The auxiliary matrix of the above considered discretized volume V,
has the following property: \textbf{X}\underline w=0. The solution
\underline{F}$(\xi)$ of eigenvalue problem (\ref{eq1}) has the
following property: either \underline{F}$(\xi)$=$\Phi
(\xi)\underline{w}$ where $\Phi (\xi)$ is the solution of problem
(\ref{eq1}) on the tile $t$, or $\underline{F}(\xi)$\underline{w}=0.
\end{lem}
\textit{Proof:} The first part of the statement is an immediate
consequence of the structure of the auxiliary matrix. The second
part of the statement is a particular case of Hersch's theorem
\cite{Her}. $\Box$ \par As elements in $\underline{w}$ are of
opposite sign on adjacent tiles,
$\underline{F}(\xi)=\Phi(\xi)*\underline{w}$ implies 0 values for
$\xi$ on the inner boundaries. If \textbf{v}, (the physical part of
the solution) does not vanish on internal sides, there is a
vanishing linear combination ($\underline{F}(\xi)\underline{w}$) of
solution values on respective points of different tiles.
\begin{thm}\label{T3.5}
Let $\underline{F}_{m}(\xi)$={\MQ}$_{m}${\Mv}$_{m}(\xi)$ be a
solution to Eq. (\ref{eq1}) over DiV $V_m$, for $\xi\in $t, m=1,2
with appropriate boundary conditions at the boundary of $V_m$. Let
{\MX}$_{m}$\underline{w}$_{m}$=0 for m=1,2. If there exists a matrix
{\MM} such that
{\MM}\underline{F}$_{1}(\xi)$=\underline{F}$_{2}(\xi)$, then
{\MM}$^{+}$ maps vector \underline{w}$_{2}$ into
\underline{w}$_{1}$.
\end{thm}
\textit{Proof: }Since $\underline{F}_m(\xi)$ is a nontrivial
solution, {\Mv}$_{m}(\xi)$ is not identically zero, in Lemma
\ref{L3.4},
 $\underline{F}_{m}(\xi)=\Phi (\xi)\underline{w}_{m}$ is excluded,
for $m=1,2$, we have $\underline{w}_{m} ^{+}\underline{ F}_{m}(\xi)
=0, m=1,2$. Since
{\MM}$\underline{F}_{1}(\xi)=\underline{F}_{2}(\xi)$, and
$\underline{w}_{1}^{+}\underline{F}_{1}(\xi) =0$, we have
{\MM}$\underline{w}_{1}^{+}\underline{F}_{1}(\xi) =0$. On the other
hand, $\underline{w}_{2}^{+}\underline{F}_{2}(\xi)
=\underline{w}_{2} ^{+}{\MM}\underline{F}_{1}(\xi)=
0=\underline{w}_{1} ^{+}\underline{F}_{1}(\xi)$, from which
immediately follows $\underline{w}_{1}={\MM}^{+}\underline{w}_{2}$.
$\Box$
\par
Theorem \ref{T3.5}. is a strong constraint on the possible matrices
transforming the solution of one boundary value problem into the
solution of the other boundary value problem.
\subsection{Transplantation rule}
Assume that discretized volumes $V_1$ and $V_2$ are equispectral.
What is the connection between the solutions $\Phi_1(x)$ and
$\Phi_2(x)$ or, in vector form, between $\underline{F}_1(\xi)$ and
$\underline{F}_2(\xi)$? In accordance with Theorem \ref{L3.2}, we
may put the eigenfunctions into vector forms:
$\Phi_m(x)=\underline{F}_m(\xi), m=1,2.$ Furthermore,
\begin{equation}\label{star}
\underline{F}_m(\xi)={\MQ}_m{\Mv}_m(\xi), m=1,2;
\end{equation}
and here
\begin{equation}\label{square}
    {\Mv}_2={\MU}{\Mv}_1.
\end{equation}
 This permits one to write
down the matrix transforming the eigenvectors into each other.
\begin{lem}[Transplantation of the solutions]\label{NewL1}
The vector forms of the solutions are connected by the following
linear transformation:
\begin{equation}\label{New-7}
    \underline{F}_2(\xi)={\MQ}_2{\MU}[{\MQ}_1^{+}{\MQ}_1]^{-1}{\MQ}_1^+
    \underline{F}_1(\xi)={\MM} \underline{F}_1(\xi).
\end{equation}
\end{lem}
Proof:  From (\ref{star}) we get
\begin{equation}\label{egy}
    {\MQ}^+_2\underline{F}_2(\xi)=\left[{\MQ}^+_2{\MQ}_2\right]{\Mv}_2(\xi)
\end{equation}
and
\begin{equation}\label{ketto}
 {\MQ}^+_1\underline{F}_1(\xi)=\left[{\MQ}^+_1{\MQ}_1\right]{\Mv}_1(\xi)
\end{equation}
Now the determination of {\MM} is straightforward:
$$\underline{F}_2(\xi)={\MQ}_2{\Mv}_2={\MQ}_2{\MU}{\Mv}_1={\MQ}_2{\MU}
\left({\MQ}^+_1{\MQ}_1\right)^{-1}{\MQ}^+_1F_1 $$ where the first
equality is the definition of $\underline{F}_2$, there we utilized
(\ref{square}), and expressed {\Mv}$_1$ from (\ref{ketto}). $\Box$
\begin{lem}
The transplantation matrix {\MM} is the solution of the following
equation:
\begin{equation}\label{New-8}
    {\MM}={\MQ}_2[{\MQ}_2^+ {\MQ}_2]^{-1} {\MQ}_2^+ {\MM}{\MQ}_1[{\MQ}_1^+ {\MQ}_1]^{-1} {\MQ}_1^+
\end{equation}
\end{lem}
{\emph{Proof:}} From (\ref{egy}) we get
${\Mv}_2=\left[{\MQ}^+_2{\MQ}_2\right]^{-1}{\MQ}^+_2\underline{F}_2,$
using the transplantation matrix $\mathbf{M}$, we get ${\Mv}_2=
\left[{\MQ}^+_2{\MQ}_2\right]^{-1}{\MQ}^+_2 {\MM} \underline{F}_1. $
Here we use the definition of $\underline{F}_1$: ${\Mv}_2=
\left[{\MQ}^+_2{\MQ}_2\right]^{-1}{\MQ}^+_2 {\MM} {\MQ}_1{\Mv}_1$.
We obtained an expression connecting $\Mv_1$ and $\Mv_2$, this is
just matrix $\MU$: ${\MU}=
\left[{\MQ}^+_2{\MQ}_2\right]^{-1}{\MQ}^+_2 {\MM} {\MQ}_1. $ Now we
substitute this expression into the second equation in(\ref{New-7})
and arrive at (\ref{New-8}). $\Box$
\begin{lem} When the transplantation matrix $\MM$ is known,
the transformation of the solutions along the internal sides is
given by
\begin{equation}\label{New-9}
    {\MU}=[{\MQ}_2^+ {\MQ}_2]^{-1}{\MQ}_2^+\MM
    {\MQ}_1.
\end{equation}
\end{lem}
\emph{Proof:} When $\underline{F}_2={\MM} \underline{F}_1$ and
$\Mv_2={\MU}{\Mv}_1$, using (\ref{egy}) we get
$\Mv_2=\left[\MQ^+_2{\MQ}_2\right]^{-1}{\MQ}^+_2\underline{F}_2.$ In
the last term we substitute
$\underline{F}_2={\MM}\underline{F}_1={\MM\MQ}_1{\Mv}_1$ and the
claim follows. $\Box$
\subsection{Constructing equispectral volumes}
The present Section is devoted to the problem of finding
equispectral volumes. Our analysis is based on \textbf{G} and
$\Gamma $, the group and graph associated with discretized volume
$V$. Let us start by the investigations initiated by Sunada
\cite{GWW}-\cite{Suna}. Let g$\in $\textbf{G} and let {\{}g{\}}
denote the conjugacy class of g in \textbf{G}. Let \textbf{G}$_{1}$,
\textbf{G}$_{2}$ $\subset $\textbf{G} subgroups in \textbf{G}. We
say (\textbf{G}, \textbf{G}$_{1}$, \textbf{G}$_{2})$ to form a
Sunada triple if the number of elements from subgroups
\textbf{G}$_{1}$ and \textbf{G}$_{2}$ are the same in every $\{g\}$.
Let $\mathbb{M}$ be a manifold.
\begin{thm}[Sunada theorem] Let $V$ be a compact
Riemannian manifold, \emph{\textbf{G}} a finite group acting on
$\mathbb{M}$ by isometries. Suppose that (\emph{\textbf{G}},
\emph{\textbf{G}}$_{1}$, \emph{\textbf{G}}$_{2})$ is a Sunada
triple, and that \emph{\textbf{G}}$_{1}$ and \emph{\textbf{G}}$_{2}$
act freely on $\mathbb{M}.$ Then the quotient manifolds
$\mathbb{M}_{1}$=\emph{\textbf{G}}$_{1}$ $\backslash \mathbb{M}$ and
$\mathbb{M}_{2}$=\emph{\textbf{G}}$_{2}$ $\backslash \mathbb{M}$ are
equispectral.
\end{thm}
Starting out from the Sunada theorem, Gordon and Webb \cite{GorW}
showed how to construct planar regions in which the Laplace operator
is equispectral.
\par
One can find discretized volumes V$_{m}$ equispectral to V$_{0}$ by
the Sunada theorem so that one searches for Sunada triples. To this
end, we have to find in the group \textbf{G}$_{0}$ associated with
V$_{0}$ subgroups of order N. The GAP program \cite{GAP} offers
means to solve that task. In Appendix A, we show such an
algorithm\footnote{ Courtesy of Dr. Erzs\'{e}bet Luk\'{a}cs of BME
Institute of Mathematics.}. The result is a coset representation of
\textbf{G}$_{0}$ in terms of subgroups \textbf{G}$_{1}$ and
\textbf{G}$_{2}$, from that the construction of V$_{1}$=V$_{0}$ and
V$_{2}$ is straightforward.
\par
Following the recipe by Gordon and Webb \cite{GorW}, we study the
graphs associated with V$_{0}$ and V$_{i}$. In that, the following
conjecture is utilized.
\begin{con}\label{C3.6}
Let $V_{1}$ and $V_{2}$ be equispectral volumes with associated
graphs $\Gamma_{1}$ and $\Gamma_{2 }$ and associated groups
\emph{\textbf{G}}$_{1}$ and \emph{\textbf{G}}$_{2}$. If graphs
$\Gamma_{1}$ and $\Gamma_{2}$ with respective edges $s_{i}$ and
$v_{i}$, are isomorphic, and the isomorphism of the edges is
$s_{i}=g_{i}v_{i}g^{-1}_{i}$, then $V_{1} \sim  V_{2}$ does not hold
unless all g$_{i}$ are automorphisms of tile t when
\emph{\textbf{G}}$_{1}$ and \emph{\textbf{G}}$_{2}$ are isomorphic
groups.
\end{con}
Buser \cite{Bus} has presented equispectral graphs which are not
isomorphic but the associated discretized volumes are not planar
ones.  Buser's graphs are derived from abstract groups and the
adjacency proposed by the coset representation exclude the
associated geometry to be planar. Buser, Conway, and Doyle
\cite{BCD} presented planar equispectral graphs, their "warped
propellers" are equivalent but if the underlying tile $t$ is not a
regular triangle, those are really equispectral. By means of the
tools of projective geometry, several further equispectral pairs
must be created.
\par
The equispectral DiVs seem to be related to isomorphic graphs and
isomorphic groups. This is formulated in conjecture \ref{C3.6}. In
Appendix B, we give a proof for a special case of that conjecture.
\section{An example}
Since isomorphic graphs contain the same number of vertices of
degrees 1, 2, and 3, one seeks equispectral discretized volumes by
specific features (length of walks, number of vertices of degree 3)
of the graph. To this end, we have analyzed $V$ composed of seven
regular triangles. There are 25 non-isomorphic graphs, the order of
the associated groups are 5040=7! for 12 graphs, 2520 for 10 graphs,
and 168 for 3 graphs. In accordance with conjecture \ref{C3.6},
equispectral volumes may exist within a family of non-isomorphic
graphs provided the tile $t$ is not symmetric. We show an example
first presented in Ref. \cite{GWW}. The order of either associated
group is 2520, the generators of the group $G_l$ assciated with the
DiV on the left are $c_l=(1,2)(5,6)$, $b_l=(3,5)(2,4)$, and
$a_l=(4,6)(5,7)$, whereas the generators of the group $G_r$
associated with the DiV on the right are $c_2=(1,2)(5,6)$,
$b_r=(3,5)(2,4)$, $a_r=(3,7)(2,6)$. The generators are related as
$c_r=c_l$, $b_r=b_l$, and $a_r=a_lb_la_l$. The associated graphs are
1-{\{}c{\}}-2-{\{}b{\}}-4-{\{}a{\}}-6-{\{}c{\}}-5(-{\{}b{\}}-3)-{\{}a{\}}-7
(left) and
7-{\{}a'{\}}-3-{\{}b'{\}}-5-{\{}c'{\}}-6-{\{}a'{\}}-2-(-{\{}c'{\}}-1)-{\{}b'{\}}-4
(on the right). (Here i-{\{}c{\}}-j stands for vertices (i,j)
connected by an edge of type c. A vertex of degree 3 has two
connecting walks, one of them is put into parentheses). The
discretized volumes shown in Fig. 4, are the so called Schreier
graphs of groups \textbf{G}$_l$ and \textbf{G}$_2$. These graphs are
associated with the discretized volumes depicted in \ref{Fig4}. The
DiV on the left/right is associated with subscript $l$/$r$,
respectively. The auxiliary matrices are
\begin{equation}\label{Exa-1}
    \emph{\MX}_l=\left(
                   \begin{array}{ccccccc}
                     1 & 1 & 0 & 0 & 0 & 0 & 0 \\
                     1 & 2 & 0 & 1 & 0 & 0 & 0 \\
                     0 & 0 & 1 & 0 & 1 & 0 & 0 \\
                     0 & 1 & 0 & 2 & 0 & 1 & 0 \\
                     0 & 0 & 1 & 0 & 3 & 1 & 1 \\
                     0 & 0 & 0 & 1 & 1 & 2 & 0 \\
                     0 & 0 & 0 & 0 & 1 & 0 & 1 \\
                   \end{array}
                 \right)
\end{equation}
and
\begin{equation}\label{Exa-2}
    \emph{\MX}_r=\left(
                   \begin{array}{ccccccc}
                     1 & 1 & 0 & 0 & 0 & 0 & 0 \\
                     1 & 3 & 0 & 1 & 0 & 1 & 0 \\
                     0 & 0 & 2 & 0 & 1 & 0 & 1 \\
                     0 & 1 & 0 & 1 & 0 & 0 & 0 \\
                     0 & 0 & 1 & 0 & 2 & 1 & 0 \\
                     0 & 1 & 0 & 0 & 1 & 2 & 0 \\
                     0 & 0 & 1 & 0 & 0 & 0 & 1 \\
                   \end{array}
                 \right).
\end{equation}
$rank(\MX_l)=rank(\MX_r)=6$, the eigenvalues are
$(0,0.225377,1,1,2.18589,3.36041,4.22833)$. The structure matrices
are
\begin{equation}\label{Exa-3}
    \emph{\MQ}_l=\left(
                   \begin{array}{cccccc}
                     0 & 0 & 0 & 0 & 0 & 1 \\
                     0 & 0 & 0 & 1 & 1 & 1 \\
                     1 & 1 & 0 & 0 & 0 & 0 \\
                     0 & 0 & 0 & 0 & 0 & 0 \\
                     0 & 1 & 1 & 0 & 0 & 0 \\
                     0 & 0 & 1 & 1 & 0 & 0 \\
                     1 & 0 & 0 & 0 & 0 & 0 \\
                   \end{array}
                 \right)
\end{equation}
and
\begin{equation}\label{Exa-4}
    \MQ_r=\left(
                   \begin{array}{cccccc}
                     0 & 0 & 0 & 0 & 0 & 1 \\
                     0 & 0 & 0 & 0 & 1 & 1 \\
                     1 & 0 & 0 & 0 & 0 & 0 \\
                     0 & 0 & 0 & 1 & 1 & 0 \\
                     1 & 1 & 1 & 0 & 0 & 0 \\
                     0 & 0 & 1 & 1 & 0 & 0 \\
                     0 & 1 & 0 & 0 & 0 & 0 \\
                   \end{array}
                 \right).
\end{equation}
The first trial is to express the solution along internal boundaries
as linear combinations of the solution along the corresponding
internal boundaries. Since solutions along the side of the same type
may be combined only, we may try{\footnote{For simplicity's sake the
normalization factor is omitted.}}
\begin{equation}\label{Exa-5}
    \MU=\left(
        \begin{array}{rrrrrr}
          0 & 0 & 0 & 1 & 0 & -1 \\
          0 & 1 & 0 & 0 & 1 & 0 \\
          1 & 0 & 1 & 0 & 0 & 0 \\
          0 & 0 & 0 & 1 & 0 & 1 \\
          -1 & 0 & 1 & 0 & 0 & 0 \\
          0 & 1 & 0 & 0 & -1 & 0 \\
        \end{array}
      \right)
\end{equation}
Using Eq. (\ref{New-7}), we get
\begin{equation}\label{Exa-6}
    \MM=\frac{3}{7} \left(
                             \begin{array}{rrrrrrr}
                               0 & 1 & 0 & -1 & 0 & -1 & 0 \\
                               1 & 0 & 0 & 1 & -1 & 0 & 0 \\
                               0 & 0 & 1 & 1 & 0 & 0 & -1 \\
                               -1 & 1 & -1 & 0 & 0 & 0 & 0 \\
                               0 & 1 & 0 & 0 & 1 & 0 & 1 \\
                               1 & 0 & 0 & 0 & 0 & 1 & -1 \\
                               0 & 0 & 1 & 0 & -1 & 1 & 0 \\
                             \end{array}
                           \right) +
    \frac{4}{7}\left(
                 \begin{array}{rrrrrrr}
                   1 & 0 & 1 & 0 & -1 & 0 & -1 \\
                   0 & 1 & -1 & 0 & 0 & -1 & 1 \\
                   -1 & 1 & 0 & 0 & 1 & -1 & 0 \\
                   0 & 0 & 0 & 1 & -1 & 1 & -1 \\
                   1 & 0 & 1 & 1 & 0 & 1 & 0 \\
                   0 & 1 & -1 & -1 & 1 & 0 & 0 \\
                   -1 & 1 & 0 & -1 & 0 & 0 & 1 \\
                 \end{array}
               \right).
\end{equation}
The two linearly independent matrix gives two variants of the
transplantation matrix \MM. We investigated if it is by chance. From
the condition that the solutions should be in $\mathbb{C}^1$, after
a long computation one obtains the following general structure of
transplantation matrix \MM:
\begin{equation}\label{Exa-7}
    \MM=\left(
        \begin{array}{rrrrrrr}
          a & b & a & -b & -a & -b & -a \\
          b & a & -a & b & -b & -a & a \\
          -a & a & b & b & a & -a & -b \\
          -b & b & -b & a & -a & a & -a \\
          a & b & a & a & b & a & b \\
          b & a & -a & -a & a & b & -b \\
          -a & a & b & -a & -b & b & a \\
        \end{array}
      \right)
\end{equation}
Here $a$ and $b$ are independent, using the choices $a=0, b=1$ and
$a=1,b=0$ one gets the two linearly independent transplantation
matrices. Gordon, Makover and Webb \cite{GorMakWeb} state that the
above decomposition corresponds to decomposing a representation of
the group \textbf{G}$_l$=\textbf{G}$_r$ by $7\times 7$ matrices into
two, nonequivalent irreducible components. The degenerate solutions
are related to
\begin{eqnarray}
  \mathbf{w}_{l} &=& (1, -1, -1, 1, 1,-1,-1) \\
  \mathbf{w}_{r} &=& (1, -1,  1, 1,-1, 1,-1).
\end{eqnarray}

%
In the analysis of discretized volumes of practical calculations,
the presented considerations may serve only as a theoretical
investigation for two reasons. The first one is the large number of
copies in a discretized volume. The algorithm presented in Appendix
A works only for N$<$10, which is really far from the practical
applications (N$>$1000). The second one is in the limited choice of
equispectral volumes: they have the same number of tiles, the
external and internal boundary types should correspond in the two
DiVs. Some of the limitations may be mitigated or eliminated in the
future.

\section{Concluding remarks}
The subject of the present work is the discretized volumes (DiVs), a
tool frequently encountered in applied mathematics. Although our
investigations have been limited to DiVs composed of a small number
of copies of a tile, the conclusions are applicable to real problems
as well. We considered the solution of an eigenvalue problem in a
DiV, in which the involved operator is linear, its symmetry include
the symmetry group of the plane $\mathbb{R}^2$, and the solution
belongs to $\mathbb{C}^1$.  The goals of the present work are:
\begin{itemize}
  \item To find algebraic descriptions for discretized volumes (DiVs) in
  order to find out whether two DiVs are equispectral or not.
  \item To find criteria for two DiVs to be equispectral.
  \item To find a transplantation recipe with which we are able to transform
  the solution from one DiV into another one.
\end{itemize}
Lemma 4.2 connects a component of the formal solution, the
structural matrix $\mathbf{Q}$ and a description of the DiV, the
auxiliary matrix $\mathbf{X}$. Theorem 4.3 gives sufficient and
necessary conditions for two DiVs to be equispectral. The theorem
uses only the structural part of the formal solution, therefore the
theorem applies to a large class of operators involved in the
eigenvalue problem. The eigenvalue connected with zero solution
along every internal sides is a degenerate case: that degenerate
problem should be discussed separately. This is done in Theorem 4.5.
The transplantation rule has two components: the transformation of
the solution along internal sides and the transformation of the
solution along the copies if the tile. The relationship among them
is established by lemmas 4.6-4.8. As to the graphs associated with a
DiV, we formulated conjecture 4.10 in the hope to attract the
attention of the experts, perhaps algebraic geometry means prove
more successful here.
\par
We achieved the following results:
\begin{enumerate}
\item Formulated an eigenvalue problem with Cauchy boundary condition for DiVs and associated algebraic
descriptions with the DiVs, viz. a graph and a group.
\item By analyzing that group and graph, one can establish relations between
eigenvalue problems on specific DiVs.
\item We improved the formal solution given in Ref. \cite{MaOr}. The improvements have lead to
Lemma \ref{L3.4} and Theorems \ref{L3.2} and \ref{T3.5}. We
formulated necessary and sufficient conditions for two DiVs to be
equispectral: when the eigenvalues of the respective auxiliary
matrices are the same then and only then are equispectral two DiVs.
\item We pointed out a surprising relationship between the
eigenvalues of the auxiliary matrix \MX and the spectrum of operator
$\mathbf{A}$ over DiV $V$. Furthermore, the degenerate eigenvalues
of $V$ are associated with the zero eigenvalue of \MX. Equispectral
DiVs $V_1$ and $V_2$ have similar auxiliary matrices.
\item Conjecture \ref{C3.6} leads to a classification of possible DiVs.
We made the classification of the DiVs composed of seven triangles,
the most frequently encountered problem in the literature
\cite{GWW}-\cite{Bus2}, \cite{MaOr}. Known equispectral DiVs confirm
our observations.
\end{enumerate}
On the other hand, our investigation addresses new problems as well.
\begin{itemize}
\item It is a question if Cayley graphs of planar
equispectral volumes are always isomorphic (Conjecture \ref{C3.6}).
\item Whether there is an algorithm to solve the problem considerably
faster than with the algorithm in Appendix A? Allows that speed for
solving practical problems (N$>$1000)?
\item Buser. Conway, Doyle, and Semmler expressed their guess:
equispectral volumes are rather scarce. Is it true for very large
DiVs (i.e. for N$\approx $10$^{5})$?
\item Can the formal solution
(\ref{eq8-1}) be generalized for DiVs with different tiles?
\end{itemize}

Discretized volumes (DiVs) offer a way in which non-equivalent
geometries can be found to solve an equation and a simple
transplantation rule can also be given. Then the transplantation is
exact. It is true that in the present form the procedure is simple
and of restricted use. However, there are reserves to be exploited.
Triangles that can be transformed into each other by a linear map
may allow for an extension of the method.
\section{Appendix A}
The GAP program given below has been written to find discretized
volumes $V_{i}$ equispectral to a given discretized volume $V_{0}$.
$V_{0}$ is defined by its three generators. The discretized volumes
$V_{i}$ are characterized by the number of copies connected by sides
of type $a, b$ and $c$. The version presented here has been made to
verify if the program finds the equivalent discretized volume pairs
discovered by Gordon and Webb, see Fig. 3.
\begin{verbatim}
# # Discretized volume V is given by the three generators of group #
G. # Now V is made up from 7 copies ot tile t # # The search starts
from V0 given also by three generators d, e and f

\# d:=(3,7)(2,6); e:=(2,4)(3,5); f:=(1,2)(5,6); {\#} LogTo("Tri7");
G:=Group(d,e,f); Size(G); CT:=[];; ri:=[];
cc:=ConjugacyClassesSubgroups(G);;
ccr:=List(cc,x-$>$Representative(x));;
ccf:=Filtered(ccr,x-$>$Size(x)*7=Size(G));; {\#} {\#} In ccf, we
have collected all subgroups of index 7 {\#} Next we apply
generators d, e, f, to them {\#} for i in [1..Size(ccf)] do
Append(CT,CosetTableBySubgroup(G,ccf[i]));Append(ri,[i]); od; {\#}
{\#} In CT, we have the discretized volumes represented by coset
tables
{\#} (One line is associated with ech generator and its
inverse)
{\#}
\end{verbatim}
\section{Appendix B: A proof for a special case of Conjecture \ref{C3.6}}
\begin{prop}\label{B.1}
A Cayley-graph is given ($\Gamma^A)$, which is depicted with
isosceles triangle nodes ($N)$, the graphic is denoted by $A$. If
the equal sides of the node are $a$ and $c$, and the third side is
denoted by $b$, then a change of $a$ and $c$ in the graph means a
reflection of the graphic: $\mathbf{T}A=B$. (The graph we get after
the change is $\Gamma^B$. The reflection is denoted by $\mathbf{T}$,
and $B$ is the graphic we get after the change in the graph.)
\end{prop}
\textit{Proof: }For the proof we use complete induction. If we start
drawing $\Gamma^A$ and $\Gamma^B$ at the same node ($N_0 )$ fixed at
a position, then the statement holds for this zeroth order case: the
node is symmetric to the altitude perpendicular to the base:
$\mathbf{T}M_0 =M_0 $. Side $a$ and $c$ are mirror images of each
other, side $b$ is reflected to itself.

Let us denote the sides of triangle $N_i $ by $a_i ,b_i ,c_i $.
Thus: $\mathbf{T}a_0^A =c_0^B ,\mathbf{T}b_0^A =b_0^B
,\mathbf{T}c_0^A =a_0^B $.

Then, in each step a new triangle ($N_i )$ is obtained and drawn in
both graphics by reflecting a formerly got triangle. We show, that
if the statement is true for step $n$ (that means $\mathbf{T}A_n
=B_n )$, then it holds for step ($n+1)$ (meaning $\mathbf{T}A_{n+1}
=B_{n+1} )$. We distinguish two cases:

If reflection $r$ (here $r$ denotes a reflection to a side and it
affects only one triangle, when $\mathbf{T}$ is the reflection to
the ordinary axis of symmetry and it affects the whole graphic) is
made with respect to side $b$, then we are reflecting to such sides
in the two graphics, that are mirror images of each other:
$\mathbf{T}b_i^A =b_i^B $. It also holds for the actually reflected
triangles: $\mathbf{T}N_i^A =N_i^B $ (as $\mathbf{T}A_i =B_i  \quad
\forall i\le n$-re). Using Lemma \ref{LB2}. the statement concludes.
It is subsequently true, that: $\mathbf{T}a_i^A =c_i^B
,\mathbf{T}b_i^A =b_i^B ,\mathbf{T}c_i^A =a_i^B $.

If the reflection is made with respect to side $a$ or $c$, then
beside $\mathbf{T}N_i^A =N_i^B $ we can also observe that
$\mathbf{T}a_i =c_i $ and $\mathbf{T}c_i =a_i $, so Lemma \ref{LB2}.
can be used as well. So: $\mathbf{T}a_i^A =c_i^B ,\mathbf{T}b_i^A
=b_i^B ,\mathbf{T}c_i^A =a_i^B \quad \forall i$.
\begin{lem}\label{LB2}
Let us consider a reflection with respect to a certain ($a$, $b$ or
$c)$ side ($t)$. If $\mathbf{T}N_i^A =N_i^B $ and
$\mathbf{T}r^A=r^B$ then $\mathbf{T}(rN_i^A )=rN_i^B $, where $r^A$
and $r^B$ denote the reflection axis in graphic $A$ and $B$
respectively.
\end{lem}
\textit{Proof:} the reflection of a triangle can be described with
the reflection of its vertices. Two vertices of $N_i^A $ are on the
reflection axis, so they are transformed by $\mathbf{T}$ into the
appropriate vertices in the other graphic. Third vertices are mirror
images of each other, as it can be seen from basic coordinate
geometrical considerations. (Reflection $r$ should be considered as
a coordinate geometrical transformation, then it is trivial, that
$\mathbf{T}$ transforms those vertices into each other.) So:
$\mathbf{T}a_i^A =c_i^B ,\mathbf{T}b_i^A =b_i^B ,\mathbf{T}c_i^A
=a_i^B $. $\Box$
\section{Acknowledgement} The kind assistance in writing the
GAP algorithm of Dr. Erzs\'{e}bet Luk\'{a}cs is acknowledged. The
authors are indebted to Dr. Jen\"o Szirmai and Dr. Zolt\'an
Szatm\'ary for their comments.
\newpage
\begin{figure}
  \includegraphics[width=10cm]{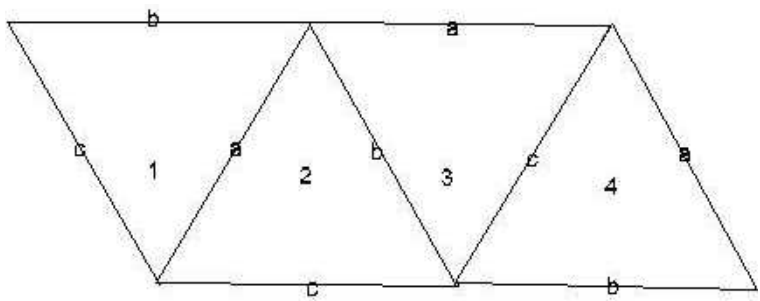}\\
  \caption{A discretized volume from four triangles\newline
  The sides are labelled by the generators}\label{Fig1}
\end{figure}
\begin{figure}
  \includegraphics[width=10cm]{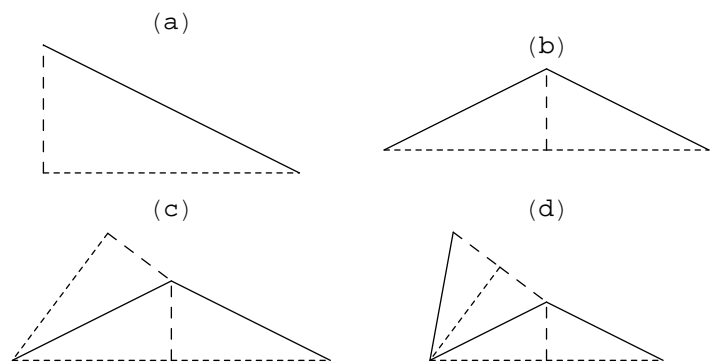}\\
  \caption{Sequence of gluing  }\label{Fig2}
\end{figure}
\begin{figure}
  \includegraphics[width=15cm]{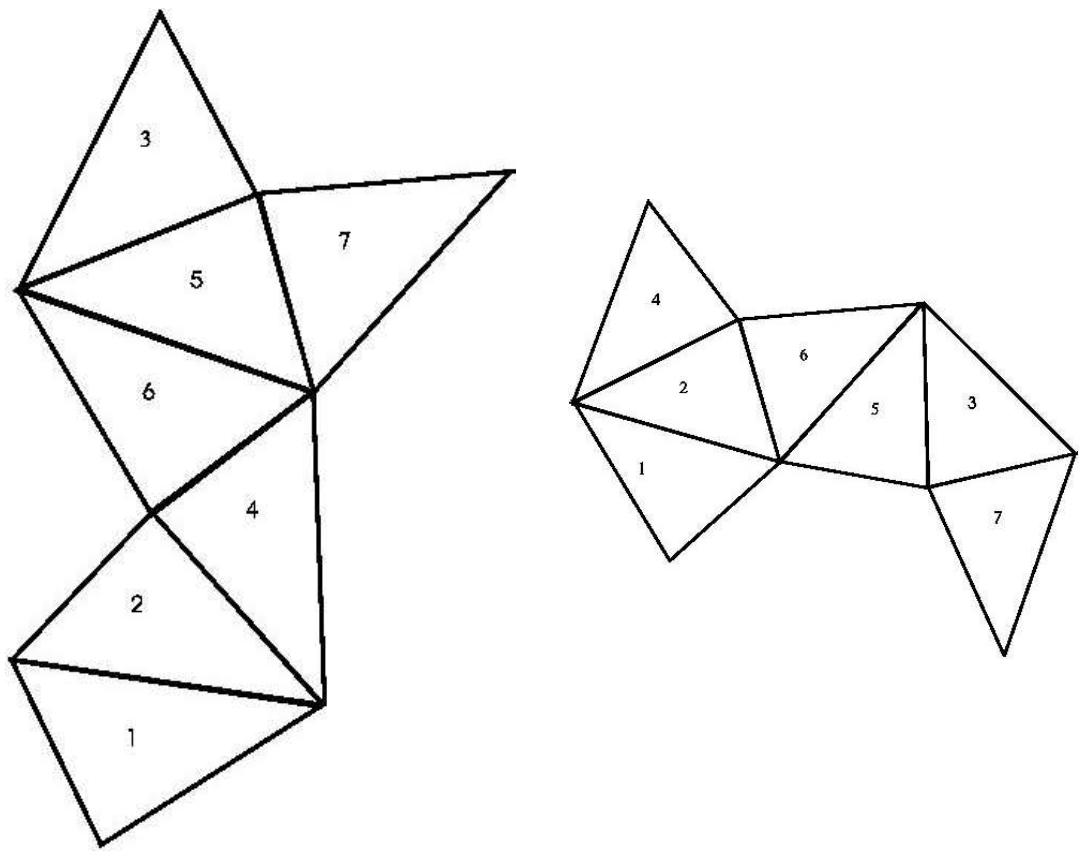}\\
  \caption{Equivalent volumes from 7 triangles }\label{Fig4}
\end{figure}
\bibliographystyle{amsplain}
\end{document}